\documentclass{article}
\usepackage{spconf,amsmath,graphicx}
\usepackage{epstopdf}
\usepackage{graphicx,epsfig,amssymb,amsmath,bm,cite,stfloats,lettrine,subfigure,booktabs}
\usepackage{color, soul}
\usepackage[linesnumbered,ruled]{algorithm2e}
\usepackage{graphics}
\usepackage{graphicx}
\usepackage[misc]{ifsym}
\usepackage{setspace}
{%
\def\theequation{\mbox{\arabic{equation}a,b}}
\begin{eqnarray}}%
{\end{eqnarray}}

\newcounter{eqnsaved}%
\title{Genetic Algorithm Optimized Support Vector Machine in
NOMA-based Satellite Networks with Imperfect CSI}
\name{Xiaojuan Yan$^{1,2}$, Kang An$^{3}$, Cheng-Xiang Wang$^{1,4,^*}$, Wei-Ping Zhu$^{5}$, Yusheng Li$^{3}$, Zhiqiang Feng$^{2}$\thanks{This work was supported by the National Key R$\&$D Program of China under grant 2018YFB1801101, the NSFC under grant 61960206006 and 61901502, the Fundamental Research Funds for the Central Universities under grant 2242019R30001, the Research Fund of National Mobile Communications Research Laboratory, Southeast University, under grant 2020B01, and the EU H2020 RISE TESTBED2 project under grant 872172.}}
\address{$^1$\small{National Mobile Commun. Research Lab, School of Info. Science and Eng., Southeast University, Nanjing 210096, China}\\
$^2$\small{College of Mechanical, Naval Architecture and Ocean Engineering, Beibu Gulf University, Qinzhou 535011,
China.}\\
$^3$\small{Sixty-third Research Institute, National University of Defense Technology, Nanjing 210007,
China}\\
$^4$\small{Purple Mountain Laboratories, Nanjing, 211111,
China}\\
$^5$\small{Department of Electrical and Computer Engineering, Concordia University, Montreal, QC H3G 1M8,
Canada.} \\
\small{Email: \{yxj9609, fzqsjtu, lys63s\}@163.com, ankang89@nudt.edu.cn, chxwang@seu.edu.cn, weiping@ece.concordia.ca.}\\
\small {$^*$}Corresponding Author: Cheng-Xiang Wang.
}

\begin{document}
%
\maketitle
\begin{abstract}
With the help of a power-domain non-orthogonal multiple access (NOMA) scheme, satellite networks can simultaneously serve multiple users within limited time/spectrum resource block. However, the existence of channel estimation errors inevitably degrade the judgment on users' channel state information (CSI) accuracy, thus affecting the user pairing processing and suppressing the superiority of the NOMA scheme. Inspired by the advantages of machine learning (ML) algorithms, we propose an improved support vector machine (SVM) scheme to reduce the inappropriate user pairing risks and enhance the performance of NOMA based satellite networks with imperfect CSI. Particularly, a genetic algorithm (GA) is employed to optimize the regularization and kernel parameters of the SVM, which effectively improves the classification accuracy of the proposed scheme. Simulations are provided to demonstrate that the performance of the proposed method is better than that with random user paring strategy, especially in the scenario with a large number of users.
\end{abstract}
\section{Introduction}
\label{sec:intro}
Benefiting from the ubiquitous coverage and emergency broadcast provision, satellite networks are expected to not only effectively offload traffic for terrestrial networks, but also provide satisfactory services for rural/harsh areas where deployments of terrestrial components are economically unfavorable in the upcoming 5G era \cite{Lamp1}, \cite{Lamp2}. However, conventional orthogonal multiple access (OMA) scheme, such as time/frequency/code-division multiple access, largely limit the improvements in spectrum utilization efficiency and the number of served users\cite{Lamp3}. In this regard, 
research efforts have been devoted to the investigation of power domain non-orthogonal multiple access (NOMA) scheme, which can simultaneously serve multiple users by superposing signals in power domain at transmitter and using successive interference cancellation (SIC) strategy at receiver \cite{C1}.
Until now, several works have been found in various satellite architectures. In GEO/MEO/LEO satellite networks from the perspective of resource utilization efficiency and transmission rate [4],\cite{Lamp5}. In hybrid satellite terrestrial networks to enhance the transmission reliability of users \cite{Lamp6} and \cite{Lamp7}. In unmanned aerial vehicles with min-max outage probability of vehicles \cite{Lamp8}.

Although the aforementioned works [3-8] have verified the superiority and feasibility of the NOMA scheme in satellite networks, they are restricted to the scenarios with perfect knowledge of channel state information (CSI) and proper user pairing to ensure significant differences in NOMA users' link conditions. In practice, however, perfect CSI may be unavailable due to delay, path loss, and user mobility for satellite networks \cite{Lamp9}. As a result, inevitable imperfect CSI can make the judgment on users' order inaccurate, disrupt decoding order, and degrade the NOMA scheme \cite{C10},\cite{C11}. Under this condition, it is vital to geer towards practical NOMA based satellite networks where imperfect CSI is taken into account, to meet the urgent requirement of improving the resource utilization efficiency of satellite networks in the 5G era \cite{Lamp12}.

Recently, machine learning (ML) algorithms, which can exploit prior knowledge from given scenarios and further function to identify unknown or similar cases, have been incorporated in various fields for the purpose of design and performance evaluation of telemetry, detection, and communication paradigms \cite{Lamp13},\cite{Lamp14},\cite{Lamp15}. Motivated by these benefits, here we resort to ML algorithms to effectively classify users into different categories, which can achieve an efficient and reliable user pairing and ensure a proper decoding order. Particularly, in the considered NOMA-based downlink satellite networks with imperfect CSI, we use a genetic algorithm (GA) to optimize the penalty and kernel parameters of a support vector machine (SVM) scheme, to effectively improve the classification accuracy of the SVM and ensure the link difference in different categories. Simulations are finally provided to show that the superior performance of the proposed method as compared to that with random user paring strategy, especially in the scenario with a large number of users.

\section{System Model}
\label{sec:format}
Consider a downlink NOMA--based satellite system, that is designed to serve $M$ ($M\ge2$) users with the help of the NOMA scheme. These $M$ users are randomly deployed in area A with different channel statistical prosperities and channel estimation errors. Before introducing the proposed user pairing strategy, the link model and spatial distribution model of those $M$ users, and the signal model using the NOMA scheme are first introduced as follows:
\subsection{Link model}
\label{ssec:subhead}
Although different fading channel models, i.e., Loo and Karasawa, have been proposed for satellite links, the Shadowed-Rician (SR) model has been widely used in various fixed/mobile scenarios for a variety of frequency bands\cite{Lamp16}, such as S-band, L-band and Ka-band, due to its nice mathematical properties \cite{Lamp17},\cite{Lamp18}. In the SR fading model, the channel coefficient for the link of satellite$\to$User ${I }$ is described as\cite{Lamp19}
$
h_I = \bar h_I + \tilde h_I,
$
where $\bar h_I$ and $\tilde h_I$ represent the amplitudes of the line of sight (LOS) component and scatter component, respectively. As proposed in [16], $\bar h_I$ follows a Nakagami distribution with an average power ${\Omega _I}$ and a fading parameter ${{m_I}}$ $\left( {{m_I }> 0} \right)$, while $\tilde h_I$ follows a Rayleigh distribution with an average power of $2{b_I}$.
Moreover, the authors in [16] have also associated parameters $b _I$, $m _I$, and $\Omega _I$ to elevation angle $\theta _I $, within the range $20^0\le \theta _I \le 80^0$, as
\begin{eqnarray}\label{eq13}
b_I \left( {\theta _I } \right) &&\!\!\!\!\!\!\!\!\!=  - 4.7943 \times 10^{ - 8} \theta _I^3  + 5.5784 \times 10^{ - 6} \theta _I^2\nonumber\\&& - 2.1344 \times 10^{ - 4} \theta _I^{}  + 3.271 \times 10^{ - 2},\nonumber\\
m_I \left( {\theta _I } \right) &&\!\!\!\!\!\!\!\!\!= 6.3739 \times 10^{ - 5} \theta _I^3  + 5.8533 \times 10^{ - 4} \theta _I^2 \nonumber\\&& - 1.5973 \times 10^{ - 1} \theta _I^{}  + 3.5156,\nonumber\\
\Omega _I \left( {\theta _I } \right) &&\!\!\!\!\!\!\!\!\!= 1.4428 \times 10^{ - 5} \theta _I^3  - 2.3798 \times 10^{ - 3} \theta _I^2 \nonumber\\&& + 1.2702 \times 10^{ - 1} \theta _I^{}  - 1.4864.
\end{eqnarray}
\subsection{Spatial distribution model}
By approximating network nodes as spatial point processes, various mathematical models, have been proposed to model the locations of wireless nodes. Among those models, binomial and Poisson point process (PPP) is the most popular model due to its well-studied and highly tractable nature \cite{Lamp20}. The probability of $M$ nodes with PPP in the region area $A$ of node density $\lambda$ can be expressed as $P\left( M \right) = \frac{{\lambda A}}{{M!}}e^{ - \lambda A}$.

\subsection{Signal model}
\label{ssec:subhead}
Using the NOMA scheme, the satellite transmits
a superposed signal $x$, i.e., $x = \sqrt {\alpha P_s } x_n  + \sqrt {\left( {1 - \alpha } \right)P_s } x_f$, where $x_n$ and $x_f$ denote the unit signals of User ${n }$ and User $f$, which are selected either with a random strategy as proposed in \cite{Lamp6} and \cite{Lamp7} or a new strategy that we will propose in this paper, and $\alpha$ is the split factor of transmission power $P_s$ to User ${n }$. At the receiver side of User ${f } $, which is assumed to experience bad link quality, the received signal is
\begin{equation}\label{4}
y_f=  h _f x+n_f,
\end{equation}
where $n_m$ $(m\in (n,f ))$ is the noise at at User ${m } $ with zero mean and $\delta_{m}^2$ variance.
However, due to the effect of delay, path loss, and mobility, the CSI estimated at User $m$ using maximum likelihood or least square estimator \cite{Lamp21},\cite{Lamp22} would be imperfect and can be given by
$
\hat h _m  = h_m  + e_m,
$
where
$\hat h _m$ and $e_m$ denoting the estimated channel coefficient and estimated channel error with $e_m~{\mathcal {CN}}(0,\delta_{m}^2/L)$, respectively \cite{Lamp22}, with $L$ being the number of training symbols. Since that $\hat h _m$ and $e_m$ are orthogonal, and according to the principle of downlink NOMA, user with the worst channel quality decodes its own information directly.
Thus, the signal-to-interference-plus-noise ratio (SINR) at User ${f } $ is

\begin{equation}\label{eq4}
\gamma _f^{\rm{}}  = \frac{{\left( {1 - \alpha } \right)P_s h _f^2}}{{\alpha P_s h _f^2+ \delta_{f}^2/L+\delta_{f}^2 }},
\end{equation}\label{eq4}At the same time, User ${n} $, which is assumed to experience good link quality, decodes and subtracts $x_f$ firstly by using the SIC strategy, and then, detects its own information, $x_n$. Thus, with imperfect CSI, the decoding and the received SINR at User ${n } $ can be, respectively, obtained as
\begin{equation}\label{eq6}
\gamma _{n \to f}^{\rm{}} = \frac{{\left( {1 - \alpha } \right)P_s h _n^2}}{{\alpha P_s  h_n^2+ \delta_{n}^2/L+\delta_{n}^2 }},
\end{equation}\label{eq6}
\begin{equation}\label{eq7}
\gamma _n^{\rm{}}  = \frac{{{\alpha } P_s h_n^2}}{{ \delta_{n}^2/L+\delta_{n}^2 }}.
\end{equation}\label{eq7}

In light of the above discussions, we see that imperfect CSI inevitably introduces extra interference, and even worse, it may make the judgment on users' CSI inaccurate, disrupt the decoding order, and degrade the superiority of the NOMA scheme. Under this consideration, the aim of this paper is to use the GA optimized SVM strategy to classify the users into certain types properly and achieve an efficient and reliable user pairing for proper decoding order.

\section{GA optimized SVM for user pairing}
\label{sec:pagestyle}

In this section, we propose a user pairing strategy based on the GA optimized SVM.
\subsection{SVM}
\label{ssec:subhead}
As an optimization algorithm based on the principle of structural risk minimization, SVM is able to learn the characteristics of different categories from training users and function a classifier or  regressor to identify unknown or similar testing users \cite{Lamp23},\cite{Lamp24}. In this paper, the SVM strategy is used to classify a satellite user into far users or near users. Specially, we assume that $N$ $(N\le M-2)$ satellite users are randomly selected from those $M$ users to form the set of training examples and others form the testing set. Then, with the help of the SVM, these training examples can be reasonably separated by a hyperplane $g({\bf{x}})$, which has the largest distance to the nearest training--data point of any class, and this kind of optimization problem can be written as
\begin{subequations}
\begin{align}
\mathop {\min }\limits_{{\bf{w}},\zeta _i ,C}~~~~&C \sum\limits_{i = 1}^N {{ \zeta _i } }  + \frac{1}{2}\left\| {\bf{w}} \right\|^2,\\
s.~t.,~~~~&y_i\left( {\left\langle {{\bf{w}},\Phi \left( {{\bf{x}}_{{i}} } \right)} \right\rangle  + b} \right)  \ge1- \zeta _i^{},\\
&\zeta _i^ {}   \ge 0,   i = 1,2, \cdots ,N,
\end{align}
\end{subequations}\label{eq7}where regularization parameter $C$ represents the tradeoff between increasing the margin-size and ensuring that the ${{\bf{x}}_{{i}} }$ lies on the correct side of the margin, the slack variable $\zeta _i$ controls how far a point lies on the wrong side of $g({\bf{x}})$,
${\bf{w}}$ is the normal vector of $g({\bf{x}})$, ${{\bf{x}}_{{i}} }$ is the
two-dimensional link budget of User $i$, $\Phi {\left( \cdot \right)}$ is the transformation function (mapping ${{\bf{x}}_{{i}} }$ to a feature space), $\left\langle{\cdot}\right\rangle$ represents the dot product, $y_i$ is the $i$th label (i.e., $-1$ stands for far users and $+1$ for near users).
With the help of the Lagrangian function, we have
\begin{eqnarray}\label{eq13}
L\left( {{\bf{w}},b,\zeta _i ,a,r} \right)\!\!\!\!&=&\!\!\!\!\!\!\frac{1}{2}\left\| {\bf{w}} \right\|^2  + C\sum\limits_{i = 1}^N {\zeta _i } - \sum\limits_{i = 1}^N {r_i } \zeta _i^{} \nonumber\\\!\!\!\!\!\!\!\!&-& \!\!\!\!\!\!\!\sum\limits_{i = 1}^N {a_i } \left( {y_i \left( {\left\langle {{\bf{w}},\Phi \left( {{\bf{x}}_{{i}} } \right)} \right\rangle  \!+ \!b} \right)\! - \!1 \!+\! \zeta _i^{} } \right).
\end{eqnarray}\label{eq13}After differentiating the above equation with respect to ${\bf{w}}$,  $b$, and  $\zeta _i $, the dual problem of (6) then becomes
\begin{subequations}
\begin{align}
\mathop {\max }\limits_a ~~~~&\sum\limits_{i = 1}^N {a_i }  - \frac{1}{2}\sum\limits_{i,j = 1}^N {a_i a_j } y_i y_j {\rm K} \left( {{\bf{x}}_i ,{\bf{x}}_j } \right),\\
 s.~t.,~~~~&0 \le a_i  \le C,  i = 1,2, \cdots ,N, \\&\sum\limits_{i = 1}^N {a_i } y_i  = 0,
\end{align}
\end{subequations}
where ${\rm K} \left( {{\bf{x}}_i,{\bf{x}}_j } \right)=\Phi \left( {{\bf{x}}_i }\right)^T\Phi \left( {{\bf{x}}_j }\right)$ denotes a kernel function. Here, we adopt a widely used Gaussian kernels ${\rm K}\left( {x,z} \right) = \exp ^{} \left( { - \frac{{\left\| {x - z} \right\|^2 }}{{2\delta ^2 }}} \right)$ with $\delta$ being the width parameter. By using the quadratic programming algorithms, the optimal $a_{}^ * $ in (8) can be written as
$
 a_{}^ *   = \left( {a_1^ *  ,a_2^ *  , \cdots ,a_N^ *  } \right)^T,
 a_{i}^ *\ge 0.
$
 Then, we have
$
   {\bf{w}}^ *   = \sum\limits_{i = 1}^N {a_i^ *  y_i } {\bf{x}}_i,
$
 and
$
b^ *   = y_j  - \sum\limits_{i = 1}^N {a_i^ *  y_i } {\rm K}\left( {{\bf{x}}_i ,{\bf{x}}_j } \right).
$
Finally, the hyperplane $g({\bf{x}})$ can be given by
$
g\left( {\bf{x}} \right) = {\bf{w}}^ *  {\bf{x}} + b^ * .
$

\subsection{GA optimized SVM}
\label{ssec:subhead}
From above analysis, we can find that the value of $a_{}^ * $, which has a direct impact on the hyperplane $g({\bf{x}})$, is closely related to the values of $C$ and $\delta$. This is because a small $C$ value allows $\zeta _i$ to be a large data and means a wide margin, while a large $C$ value means a narrow margin. Furthermore, if $\delta$ is overestimated, the Gaussian kernel will behave almost linearly and lose its non-linear property. If underestimated, on the other hand, Gaussian kernel will lack regularization and sensitivity to noise in training data.

\begin{algorithm}
\caption{GA optimized SVM training algorithm.}
\KwIn{Link budgets, such as $\theta _i$ and $e_i$ ($i = 1,2, \cdots ,N$), and labels of $N$ training users.}
\ Initialize the population size, crossover probability, mutation probability, and max number of iterations as: $Z=20$, $p_x=0.6$, $p_f=0.1$, and $g_{max}=300$;\\
\ Generate $Z$ initial $pop$s randomly and set $z=g=1$;\\
\For{$g \le g_{max}$}
{
\If{$z\le Z$}
{
Substitute $pop_z$ into the hyperplane $g({\bf{x}})$ to obtain the related $g({\bf{x}})$, with which the mean square error (MSE) of those $N$ training users are obtained;\\
$z=z+1$;\\
}
Sort $pop$s according to their corresponding ${\rm{MSE}}$s;\\
\While{{\rm{BestMSE}}$\ge0.01$}
{
\ Select $pop$s with high fitnesses;\\
 \ Do genetic operators, such as crossover and mutation, to generate $M$ new $pop$s;\\
}
$g=g+1$\;
}
return $pop$ with the best MSE\;
\KwOut{The optimized hyperplane $g(\bf{x})$.}
\end{algorithm}

Despite the advantages of the SVM scheme, these above discussions have shown that parameters $C$ and $\delta$ can largely influence the accuracy of the SVM scheme. To choose parameters $C$ and $\delta$ smartly, in this paper, we introduce the GA algorithm to get a better combination of these two parameters in SVM. Since that the optimal solution of the GA algorithm is achieved by a series of iterative computations, here, we use a population, $pop$, which is binary encoded with length $2\times10$, to jointly represent the values of these two parameters. The main steps of GA optimized SVM training algorithm is given in ${\textbf{Algorithm 1}}$, in which fitness function is given by $fitness=\frac{N_t}{N_t+N_f}\times 100\%,$ where $N_t$ and $N_f$ respectively denote the number of true and false classifications.

\subsection{User pairing strategy}
\label{ssec:subhead}
With the optimized parameters $C$, $\delta$, and the hyperplane $g(\bf{x})$ obtained through ${\textbf{Algorithm 1}}$, the users in training set can be well classified. Using that hyperplane $g(\bf{x})$, users in testing set will be classified into the category of far users or near users  to ensure the link difference in different categories. Then, one from each of these two categories are chosen to form a NOMA group, and thus, reducing the risk of inappropriate user pairing stemmed from random selection strategy, i.e., users with similar channel qualities are chosen.

\section{Simulation Results and Analysis}
\label{sec:typestyle}

In this section, simulation results are provided to evaluate the performance of the proposed strategy. Specially, we set $L=7$, and $\delta_{n}^2=\delta_{f}^2=1$. Moreover, we assume that 50 users with various $\theta $, which randomly distribute within the area $[20^{\rm{o}}, 80^{\rm{o}}]$, are deployed within a beam spot with radius of 100 $\rm{Km}$. Besides, $N=40$ users are randomly selected to form the training set, and other users form the testing set.
\begin{figure}[!h]
\centering
  \subfigure[]{
    \label{fig:subfig:a} 
    \includegraphics[width=4.1cm,height=4.1cm]{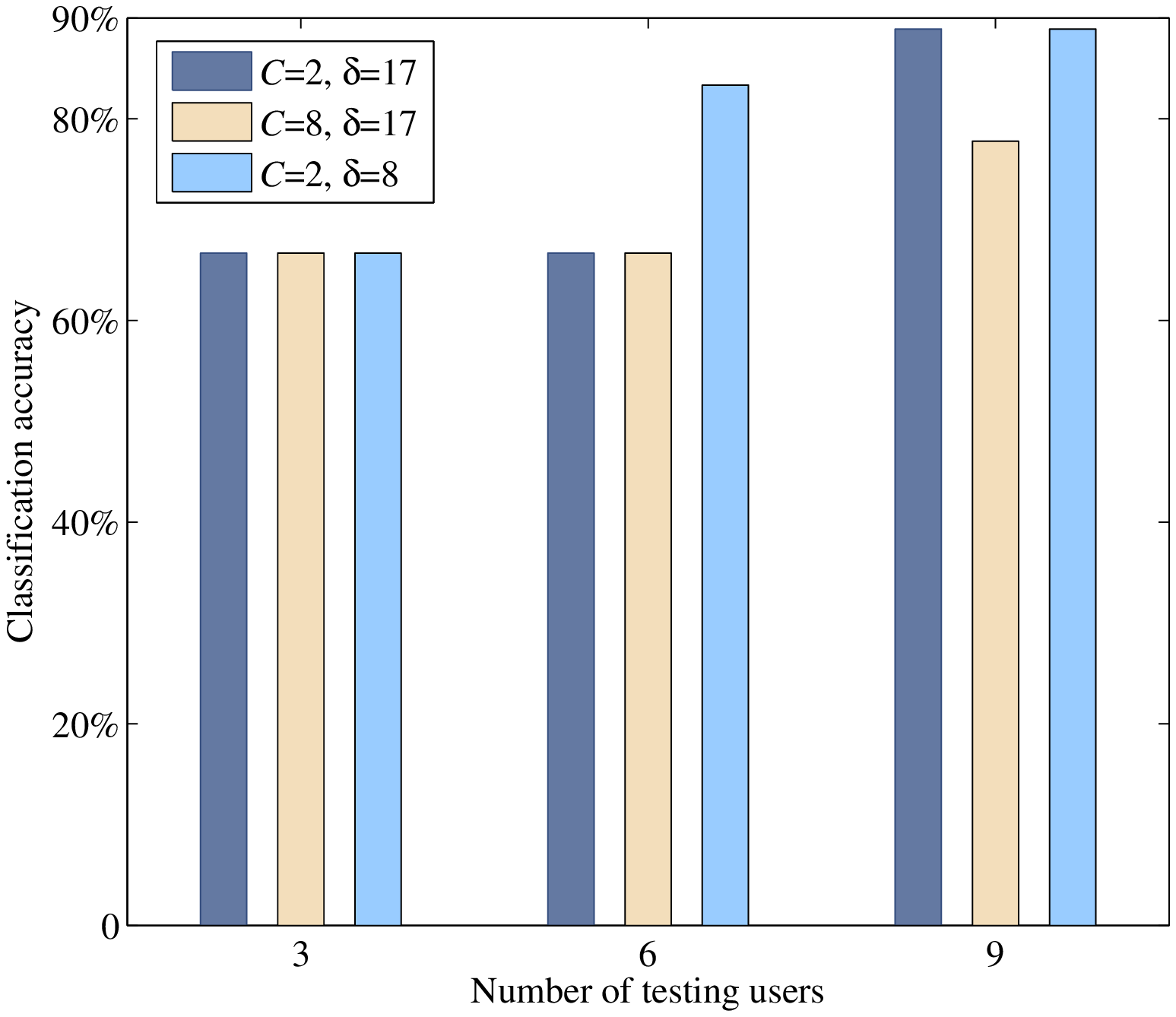}}
  \hspace{0in}
  \subfigure[]{
    \label{fig:subfig:b} 
    \includegraphics[width=4.1cm,height=4.1cm]{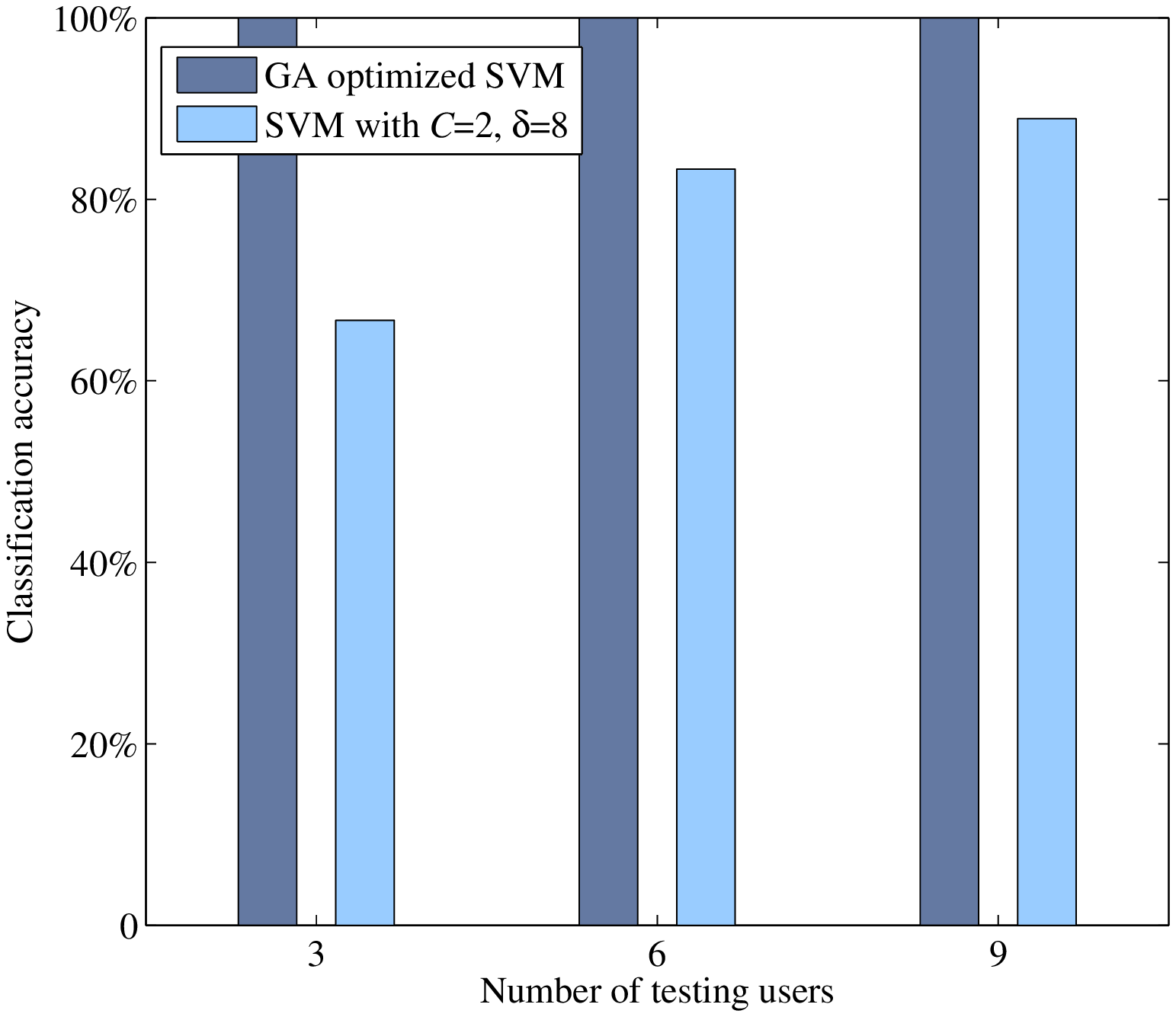}}
  \caption{Testing accuracy versus number of testing users: (a) SVM scheme with various parameters and (b) SVM and the GA optimized SVM schemes.}
  \label{fig:subfig} 
\end{figure}
\begin{figure}[!h]
\centerline{\epsfxsize=6.8cm\epsffile{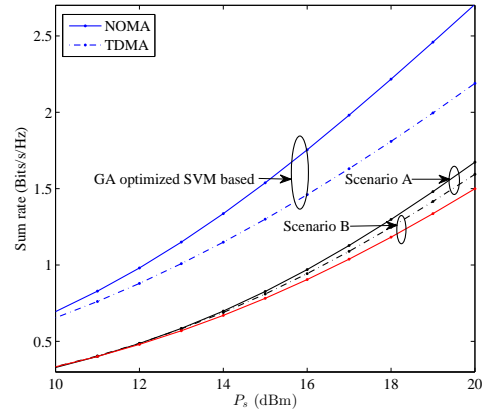}}
\caption{Sum rate versus transmission power $P_s$.}
\end{figure}

To show the effects of $C$ and $\delta$ on the hyperplane $g(\bf{x})$ of SVM classifier, we firstly plot the classification accuracy of SVM classifier with various parameter configurations for testing users in Fig.~1(a). As observed, the classification performance achieved by a small penalty parameter $C$ or a small kernel parameter $\delta$, i.e., $C=2$ or $\delta=8$, is always not worse than those achieved by a large $C$ or $\delta$ in all cases. This is because small values of $C$ and $\delta$ allow a large margin size and non-linear, respectively, for the hyperplane $g(\bf{x})$, which is built according to the link vectors of training users. Moreover, we observe that the classification accuracy varies for different numbers of testing users. Even so, the performance of SVM classifier with $C=2$ and $\delta=8$ is optimal in all cases. 

Fig.~1(b) plots the comparison of classification accuracy with the SVM and the GA optimized SVM schemes. As can be seen, the classification accuracy with the GA optimized SVM is all superior to those with the SVM. This performance improvement is achieved by using the benefit of GA strategy to select proper parameters $C$ and $\delta$, whose searching areas are correspondingly set as $[0.1, 15]$ and $[0.1,30]$. 

Finally, we compare the achievable sum rate between the proposed GA optimized SVM approach and the random selection strategy proposed in [6] and [7]. Specifically, we consider the number of testing users is $9$ and power allocation factor $\alpha=0.2$. Moreover, scenario A here means that NOMA users are selected with the random selection strategy suggested in [6], [7] with correct decoding order, while scenario B means the corresponding scenario with incorrect decoding order. As shown in Fig.~2, the sum rate performance with the NOMA scheme outperforms that with the TDMA scheme in the case of correct decoding order. Moreover, we can clearly find the superiority of the proposed strategy obviously overs scenario A. This is because with the proposed strategy, the link difference of users can be ensured, while there is marginal link difference between users in scenario A. Meanwhile, as shown in scenario B, the disrupt decoding order caused by imperfect CSI significantly degrades the NOMA scheme. This phenomenon also proves the applicability of NOMA scheme with the proposed method.

\section{Conclusions}
\label{sec:majhead}

In this paper, we have studied the application of a GA optimized SVM scheme in NOMA-based downlink satellite networks. Specifically, we selected NOMA users according to the result of a SVM classifier, whose key parameters have been optimized by using the GA. Simulation results have shown that the proposed GA optimized SVM scheme provides an effective user classification in NOMA based satellite systems. In addition, our results have also revealed the superiority and applicability of the NOMA scheme using our proposed classification strategy over the existing NOMA systems with random selection strategy.

\bibliographystyle{IEEEbib}
\bibliography{strings,refs}

\end{document}